\documentclass[rnote]{aa}

\usepackage{txfonts}
\usepackage{graphicx}
\usepackage{threeparttable}
\usepackage{natbib}

\title{Direct imaging constraints on planet populations detected by microlensing}
\author{Sascha P. Quanz\inst{1} 
\and David Lafreni\`ere\inst{2} 
\and Michael R. Meyer\inst{1}
\and Maddalena M. Reggiani\inst{1}
\and Esther Buenzli\inst{3}}
\institute{Institute for Astronomy, ETH Zurich, Wolfgang-Pauli-Strasse 27, CH-8093 Zurich, Switzerland
\and D\'epartement de Physique, Universit\'e de Montr\'eal, C.P. 6128, Succ. Centre-Ville, Montr\'eal, QC, H3C 3J7, Canada
\and  Steward Observatory, University of Arizona, Tucson, USA}
\date {Accepted: March 15, 2012 }
\begin{document}
\abstract{Results from gravitational microlensing suggested the existence of a large population of free-floating planetary mass objects. The main conclusion from this work was partly based on constraints from a direct imaging survey. This survey determined upper limits for the frequency of stars that harbor giant exoplanets at large orbital separations.}{We want to verify to what extent upper limits from direct imaging do indeed constrain the microlensing results.}{We examine the current derivation of the upper limits used in the microlensing study and re-analyze the data from the corresponding imaging survey. We focus on the mass and semi-major axis ranges that are most relevant in context of the microlensing results. We also consider new results from a recent M-dwarf imaging survey as these objects are typically the host stars for planets detected by microlensing.}{We find that the upper limits currently applied in context of the microlensing results are probably underestimated. This means that a larger fraction of stars than assumed may harbor gas giant planets at larger orbital separations. Also, the way the upper limit is currently used to estimate the fraction of free-floating objects is not strictly correct. If the planetary surface density of giant planets around M-dwarfs is described as $d f_{\rm Planet}\propto  a^\beta da$, we find that $\beta\lesssim 0.5 - 0.6$ is consistent with results from different observational studies probing semi-major axes between $\sim$0.03 -- 30 AU.}{Having a higher upper limit on the fraction of stars that may have gas giant planets at orbital separations probed by the microlensing data implies that more of the planets detected in the microlensing study are potentially bound to stars rather than free-floating. The current observational data are consistent with a rising planetary surface density for giant exoplanets around M-dwarfs out to 
$\sim$ 30 AU. Future direct imaging surveys will show out to what semi-major axis the above mentioned range of $\beta$ is valid and what fraction of the planetary mass objects detected by microlensing are indeed bound.}


\authorrunning{Quanz et al.}
\maketitle
\section{Introduction}
Gravitational microlensing is a powerful tool to study the initial mass function (IMF) of stars and sub-stellar objects for stellar populations probed along a particular line of sight. It can also effectively probe the companion mass ratio distribution (CMRD) all the way down to terrestrial mass objects. In particular for exoplanet studies the microlensing technique provided some unique results in the last years, such as the detection of a cool Super-Earth orbiting an M-dwarf at $\sim$3 AU \citep{beaulieu2006} and unprecedented constraints on the frequency of cold Neptune-like and Jupiter-like planets \citep{sumi2010,gould2010}. 

\citet{sumi2011} reported the detection of an unbound or distantly orbiting planetary mass population based on gravitational microlensing results. 
These pioneering observations could be interpreted as the low-mass end of the IMF, as bound planetary mass companions, or as ejected planets. The authors rely on constraints from a direct imaging campaign \citep{lafreniere2007} to conclude that at least 75\% of these objects are not bound to any star and were probably ejected from planetary systems via dynamical interactions. Here we show that in fact the constraints available from direct imaging are less stringent than assumed to support the claim that most of the detected objects are indeed unbound. We summarize the analyses from \citet{sumi2011} in section 2, re-analyze the relevant direct imaging data in sections 3 and 4, and discuss the implications of the re-analysis in section 5.

\section{A large population of unbound planets - derivation of current constraints}
Using the datasets from the OGLE and MOA microlensing collaborations \citet{sumi2011} identified 10 single microlensing events that lasted only a comparatively short amount of time. As these short events were not accompanied by a longer signal on which the short event was overlaid the authors concluded that the short events were caused by individual very-low mass objects that did not show any indications for a host star. Taking into account the detection efficiency of different types of microlensing events \citet{sumi2011} modeled the short events with a population of Jupiter-mass objects that are 1.8$^{+1.7}_{-0.8}$ times more frequent than main-sequence stars with masses $<$1.0 M$_{\sun}$. The masses of the objects were, however, not directly constrained by the observations and also the existence of host stars could only be ruled out to a certain separation. The mean minimum separation between the 10 inferred planets and their potential host stars is $\langle a_{\rm min}\rangle\sim$16 AU but we note that only 2 objects have minimum separations exceeding this mean value with $\sim$21 AU and $\sim$45 AU, respectively\footnote{According to \citet{sumi2011} the minimum separation in AU can be computed from the minimum separation in units of the Einstein radii of the possible host star $d_{\rm min}$(R$_{E^*}$) via $a_{\rm min}\approx 3 {\rm AU} \cdot d_{\rm min}$(R$_{E^*}$).}. In order to assess how many of these planetary mass objects might be bound to a star \citet{sumi2011} use the results from the Gemini Deep Planet Survey \citep[GDPS;][]{lafreniere2007} that aimed at the detection of planets through direct imaging. 
As explained in their supplementary material, \citet{sumi2011} estimated the upper limit for Jupiter-mass planets with separations between 10 -- 500 AU by taking mean of the 1 Jupiter-mass curve from Fig. 10 of \citet{lafreniere2007}, which leads to $f_{max}< 40\%$ (95\% confidence).
As the objects that they detected have minimum separations from their host stars within this range, \citet{sumi2011} conclude that, on average, a star hosts up to 0.4 planets, and hence at least 75\% of their objects  are unbound or free-floating planets\footnote{\citet{sumi2011} detected 1.8 planets per star from which they subtracted 0.4 to correct for those planet planets that could be bound. Hence, 1.4/1.8$\approx$75\% is the fraction of unbound planets in their sample.}.
\section{Re-analysis of Gemini Deep Planet Survey data}
The results presented by \cite{sumi2011} imply more than just a large population of unbound planetary mass objects. If their interpretation is correct, i.e., that most of the unbound planets were dynamically ejected from their planetary systems, this would also require that at least one \emph{additional} massive planet remained in each of those systems implying that, on average, $\sim$2.8 massive planets formed around each star. As this result would have significant implications on planet formation theory and the overall census of exoplanets in the Milky Way, we take a closer look at the relevant analysis. We emphasize that we do not re-analyze the suggested abundance of planetary mass objects relative to main sequence stars, i.e., 1.8$^{+1.7}_{-0.8}$. We rather focus on the constraints that can be derived from direct imaging surveys on the 10 detected microlensing objects  and on the upper limits of stars that might harbor planetary companions. 



A  very conservative approach would  be to read off directly the upper limit for stars having a 1 M$_{\rm Jupiter}$ companion between 10 AU -- 500 AU from Fig. 10 of  \citet{lafreniere2007}. This upper limit is $f_{max}<$92\% (with 95\% confidence) and is valid for any distribution of mass and semi-major axis.
 
With access to the data from the GDPS we can directly constrain the occurrence rate of massive planets on orbital separations that appear to be the most relevant in context of the microlensing results. Using the data and the approach described in detail in \citet{lafreniere2007}, we computed a maximum planet frequency ($f_{max}$) pertaining to each of the objects detected by \citet{sumi2011}. We assumed that the planetary masses are either exactly 1 M$_{\rm Jupiter}$ or between 0.5 -- 3 M$_{\rm Jupiter}$\footnote{This mass range corresponds roughly to the 68\% confidence intervals that \citet{sumi2011} put on their assumed planetary mass function which was centered around $\sim$1 M$_{\rm Jupiter}$.} and that the objects' semi-major axes are either exactly the minimum separation $a_{\rm min}$ given by \citet{sumi2011} or, to be conservative, twice this value.  The choice of these two specific separations ensures that we have captured extreme cases, i.e., all objects at small or all object at large separations. For each of the overall four cases, we averaged the upper limits over all 10 objects. The results are summarized in Table 1.


\begin{table} 
\caption{$f_{max}$ (with 95\% confidence) for different combinations of assumed planetary masses and semi-major axes.}
\label{results}	
\centering \begin{tabular}{c|cc} \hline\hline
 & 			1 M$_{\rm Jupiter}$ & 0.5 -- 3 M$_{\rm Jupiter}$\\
\hline
$a = a_{\rm min}$ &		 78\%		&	59 \%\\
$a = 2a_{\rm min}$ & 	 49\%		&	29\%\\
\hline
\end{tabular} 
\end{table}

We found that $f_{\rm max}$ can be as high as 78\% if we assume that the mass of the microlensing planets is 1 Jupiter-mass and that the derived minimum separations from the host stars correspond to the semi-major axes of the orbits. As expected, the lowest value for $f_{\rm max}$ is obtained if the mass is assumed to be anything between 0.5 -- 3 M$_{\rm Jupiter}$ and the semi-major axes are twice the minimum separations. We note, however, that this estimate is "biased": we implicitly assumed that the planetary mass distribution is flat in those cases where we considered the mass interval between 0.5 -- 3 M$_{\rm Jupiter}$. However, radial velocity (RV) searches for exoplanets suggest that lower mass planets, which are more difficult to detect by means of direct imaging, are more common than Jupiter-like analogues \citep[e.g.,][]{mayor2011}. Thus, this upper limit is likely an under-estimate. 

\section{First constraints from direct imaging surveys of M-dwarfs}
While roughly 3/4 of the stars in the sample from \citet{lafreniere2007} were FGK stars, it is usually assumed that host stars to planetary objects detected by microlensing are M-dwarfs \citep[e.g.,][]{gould2010}. So, ideally, one wants to do a similar statistical analysis based on direct imaging results from a large survey of M-dwarf. Recently, \citet{delorme2011} published first results from a large deep imaging survey of nearby M-dwarfs. They provided the detection limits for 14 objects but did not provide a comprehensive statistical analysis of their dataset. However, we can use the provided mean detection probability of their survey for different planetary masses as a function of the separation. We restrict ourselves to separations between 10 AU -- 30 AU, i.e., out to roughly twice the mean minimum separation  $\langle a_{\rm min}\rangle$ derived from the microlensing data. Reading off the mean detection probabilities of their survey at 10 AU, 20 AU and 30 AU for objects with 0.6 M$_{\rm Jupiter}$ and 1.5 M$_{\rm Jupiter}$ and averaging them results in an overall mean detection probability of $\langle p_{\rm detect}\rangle\approx$30\%. This is a good approximation for the mean detection probability of a 1 M$_{\rm Jupiter}$ object. To be consistent with the analysis of the previous section, in particular with Table~\ref{results}, we also looked at the mean detection probability for the mass range between 0.6 -- 3.0 M$_{\rm Jupiter}$ for the same separation range, which is well approximated by the values for a 1.5 M$_{\rm Jupiter}$ object. Here we find $\langle p_{\rm detect}\rangle\approx$45\%.

As no planetary companion was detected in the survey we can ask, given these mean detection probabilities, what upper limit of stars having a companion in the above-mentioned parameter space is consistent with this null-result at the 95\% confidence level. 
Applying the same Bayesian analysis as done above for the GDPS survey and as laid out in \citet{lafreniere2007} yields upper limits of $f_{\rm max}\lesssim59\%$ and $f_{\rm max}\lesssim40\%$ for the  1 M$_{\rm Jupiter}$ case and the 0.6 -- 3.0 M$_{\rm Jupiter}$ case, respectively. Implicitly, both cases assume a flat distribution of planets in linear $a$ space and the second case furthermore assumes that the planetary mass distribution is flat as well in the given mass interval.

\section{Implications on interpretation of microlensing results}
Looking at the upper limits of stars potentially harboring planets that we have estimated in the previous sections we find that in most cases these values are higher than the 40\% used by \citet{sumi2011}. Only in those cases were we looked at a planetary mass interval $\sim$0.5 -- 3 M$_{\rm Jupiter}$, with the underlying assumption that the mass distribution is flat, we obtain upper limits between $\sim$30 -- 40\%. As already mentioned by \citet{sumi2011}, if the objects were Saturn-mass objects rather than Jupiter-mass objects, then there would basically be no constraints on their occurrence rate at all. 

It's also worth mentioning that throughout this paper we used a confidence level of 95\%. Increasing the confidence level would also increase the upper limits.

Finally, it is important to note that the statistical interpretation of the direct imaging results outlined in \citet{lafreniere2007} is binomial in the sense that a star either "does not have any planet" or "does have at least one planet". Hence, having an upper limit $f_{\rm max}$=50\% for stars that may harbor one (or more) planets does not mean that, on average, each star has 0.5 planets. The upper limits indicate the maximum fraction of stars with one or more planets. Thus,  regardless of the value chosen for the upper limit, the approach used by \citet{sumi2011} to estimate the fraction of unbound planets in their sample from the value of $f_{\rm max}$ is not strictly valid.

Based on the points detailed above we believe that \citet{sumi2011} underestimated the upper limit of stars with planets and that, furthermore, the derived upper limit can not be used to compute an average number of planets per star. In consequence, the main conclusion that 75\% of the detected microlensing planets are unbound and were probably ejected from their planetary systems is overestimated. 
On the other hand the null-result from the planet search around nearby M-dwarfs by \citet{delorme2011} suggests that it's also unlikely that all of the microlensing objects are bound to a star if they are indeed Jupiter-mass objects. If there was only one planet of 1 M$_{\rm Jupiter}$ between 10 -- 30 AU per star the probability that no object was detected around the 14 M-dwarfs is $\approx6.8\times10^{-3}$.
Overall, it seems plausible that a certain fraction of the detected planets are bound to a star, that another fraction are indeed unbound but represent the low-mass end of the IMF, and that a third fraction are also unbound but were indeed ejected dynamically from their planetary systems.

\section{Predictions for future imaging surveys}
While even more recent direct imaging surveys for planets  around solar-type stars \citep[e.g.,][]{chauvin2010} were not significantly more sensitive to planets in the relevant parameter space than the survey of \citet{lafreniere2007}, future surveys on upcoming instruments (e.g., SPHERE, GPI) will shed light on the first fraction mentioned above. Also, increasing the sample size of the M-dwarf survey by \citet{delorme2011} will be of particular interest for the microlensing community. Here we can, however, use the upper limits from the current survey and combine it with results from microlensing studies to make some testable predictions for future surveys. 

The microlensing results from \citet{gould2010} suggested that the frequency of planets between 2 -- 8 AU is $f_{\rm Planet}$=0.36$\pm$0.15. In this case, the typical mass of the host stars was $\sim$0.5 M$_{\sun}$ and the mass regime of the planets was expressed in terms of mass ratio $q$ with $-4.5 <log(q) < -2$ which would correspond to a mass range of $\sim$0.02 -- 5 M$_{\rm Jupiter}$. As explained above, \citet{cassan2012} found that the fraction of bound planets between 0.5 -- 10 AU and with masses between 0.3 -- 10 M$_{\rm Jupiter}$ is $f_{\rm Planet}$=0.17$^{+0.06}_{-0.09}$. 

Using these planet frequencies, we can describe the population of planets by
\begin{equation}
d f_{\rm Planet}=Cm^\alpha a^\beta dm\;da
\end{equation}
where $C$ is a normalization constant, $m$ is the mass of the planet, $a$ the semi-major axis, and $\alpha$ and $\beta$ are free variables. We can now find a value for $\beta$ so that the microlensing results are consistent with the upper limits that we have estimated from the M-dwarf survey in section 4.  For the power-law slope of the mass function, we use the result from \citet{cumming2008}, which in linear mass space gives $\alpha=-1.31$. This result was derived from RV searches for massive planets around FGK stars, but no equivalent statistical analysis for M-dwarfs is available. Integrating equation (1) for varying $\beta$ yields different values for the constant $C$, which in turn can then be used to compute the planet frequency for the parameter space probed by the direct imaging results, i.e., $m$ between  0.6 -- 3.0 M$_{{\rm Jupiter}}$  and $a$ between 10 -- 30 AU. 

In order to get $f_{\rm Planet}\lesssim 0.40$ over this range, we find that $\beta\lesssim 0.61$ or $\beta\lesssim 0.49$ using the results of \citet{cassan2012} or \citet{gould2010}, respectively. Given the error bars in $f_{\rm Planet}$ for those studies, both values for $\beta$ are indistinguishable. This result shows that current observations do not rule out a \emph{rising} planetary surface density for giant planets around M-dwarfs between 10 -- 30 AU in linear space. In addition, we note that a value for $\beta$ in the range of 0.5 -- 0.6 provides a direct link between the two microlensing studies mentioned above and the M-dwarf results from \citet{cumming2008}. From RV measurements, \citet{cumming2008} estimated $f_{\rm Planet}\approx 0.02$ around M-dwarfs considering planetary masses between 0.3 -- 10 M$_{{\rm Jupiter}}$ and semi-major axes between $\sim$ 0.03 -- 2.5 AU \citep[cf.][]{bonfils2011}. Using the approach outlined above and extrapolating these values to the parameter space of the microlensing studies yields consistent results. Thus, a positive power-law index $\beta\lesssim$ 0.5 -- 0.6 is currently able to describe the surface density profile for gas giant planets around M-dwarfs between $\sim$0.03 -- 30 AU.

This upper limit for $\beta$ is, however, in stark contrast to the original results from \citet{cumming2008} for solar type stars. Here, $\beta=-0.61\pm0.15$ in linear space  \citep[see,][]{heinze2010} for giant planets with periods $<$2000 days, i.e., $a\lesssim 3$  AU. If we take the expectation value $\beta=-0.61$ and extrapolate again the M-dwarf frequency from \citet{cumming2008} to the parameter space of \citet{gould2010} and \citet{cassan2012} we find that $f_{\rm Planet}\approx 0.05$ and $f_{\rm Planet} \approx 0.03$, respectively, which is not consistent with observational results within the given error bars. Taking into account the 3-$\sigma$ error bar and using $\beta=-0.16$ would result in  $f_{\rm Planet}\approx 0.11$ and $f_{\rm Planet} \approx 0.06$, again outside the error bars given by \citet{gould2010} and \citet{cassan2012}.

Future surveys with SPHERE and GPI, or an extension of the survey done by \citet{delorme2011}, will reveal the number of bound giant planets between 10 -- 30 AU around M-dwarfs, empirically constrain the power-law slope $\beta$, and test to what separation the predicted value range is valid. These surveys will provide some further hints as to what mix of bound and unbound planets is the most likely -- and less extreme -- explanation for the microlensing results presented by \citet{sumi2011}.

\begin{acknowledgements} 
We thank D. P. Bennett and T. Sumi for their prompt response to our initial inquiry about the interpretation of the microlensing results. We also thank an anonymous referee for helpful and detailed comments and suggestions.  
\end{acknowledgements}

\bibliographystyle{aa.bst}
\bibliography{mybib.bib}

\end{document}